\documentclass{appolb}
\usepackage{epsfig}

\usepackage{amsmath}
\usepackage{amsfonts}
\usepackage{amssymb}

\newcommand{\be}{\begin{equation}}
\newcommand{\ee}{\end{equation}}
\newcommand{\bea}{\begin{eqnarray}}
\newcommand{\eea}{\end{eqnarray}}
\newcommand{\nn}{\nonumber}
\newcommand{\Dirac}{\rlap{\hspace{-.8mm} \slash} D}
\newcommand{\al}{\alpha}
\newcommand{\la}{\lambda}

\def\erfc{{\mbox{erfc}}}
\def\sgn{{\mbox{sgn}}}

\def\Tr{{\mbox{Tr}}}
\def\Pf{{\mbox{Pf}}}
\def\re{{\Re\mbox{e}}}
\def\im{{\Im\mbox{m}}}

\begin{document}
\eqsec  
\title{Non-Hermitian extensions of Wishart random matrix ensembles
\thanks{Presented at the 23rd Marian Smoluchowski Symposium 
``Random Matrices, Statistical Physics and Information Theory'', Krak\'ow,
  Poland, 26-30/09/2010 }%
}
\author{Gernot Akemann
\address{Fakult\"at f\"ur Physik, Universit\"at Bielefeld,
Postfach 100131, D-33501 Bielefeld, Germany}
}
\date{18. April 2011}

\maketitle
\begin{abstract}
We briefly review the solution of three ensembles of non-Hermitian
random matrices generalizing the Wishart-Laguerre (also called chiral)
ensembles. 
These generalizations are realized as Gaussian two-matrix models, where
the complex eigenvalues of the product of the two independent
rectangular matrices are 
sought, with the matrix elements of both matrices
being either real, complex or quaternion
real.
We also present the more general case depending on a non-Hermiticity
parameter, that allows us to interpolate between the corresponding three
Hermitian Wishart ensembles with real eigenvalues and the maximally
non-Hermitian case. 
All three symmetry classes are explicitly solved for finite matrix size
$N\times M$ for all complex eigenvalue correlations functions
(and real or mixed correlations for real matrix elements).
These are given in terms of the corresponding kernels built from 
orthogonal or skew-orthogonal Laguerre polynomials in the complex plane.
We then present the corresponding three Bessel kernels in the complex
plane in the microscopic 
large-$N$ scaling limit at the origin, both at weak and strong non-Hermiticity
with $M-N\geq 0$ fixed.

\end{abstract}

\section{Introduction and motivation}

The Wishart ensemble
was the first ensemble of random matrices, introduced in the study of randomized
rectangular times series matrices $C$ in order to study the spectral
properties of its symmetric and positive covariance matrix $C^T C$. 
Later the Wigner-Dyson ensembles were formulated as models for
randomized Hamiltonians $H$ of heavy nuclei to explain some of the
spectral properties of the matrix $H$ which is Hermitian (or real
symmetric or quaternion self dual). Its non-Hermitian generalizations
were introduced and studied immediately after by  Ginibre
\cite{Ginibre}, but not so for the Wishart ensembles which had to wait
a few decades. 

In order to construct such a generalization it is useful to compare to
the Ginibre ensembles and to think of them as a two-matrix problem. By
dropping the Hermiticity constraint an independent anti-Hermitian
matrix $A$ is added to the Hermitian matrix $H$, and the now complex
eigenvalues of the {\it sum} $H+A$ of the two are studied. How can one
repeat this for Wishart matrices, where the matrix $C$ is already
without symmetry? Again one can introduce a second independent random
matrix $D$, and study this time the complex spectrum of the {\it
  product} $DC$, where $D$ is no longer the transpose (or Hermitian
conjugate) of the first matrix, $C^\dag\neq D$, but has the same
rectangular dimension. 

Interestingly the solution of this two-matrix problem did not grow out
of statistical applications, where the problem has appeared in terms
of so-called time lagged, asymmetric  covariance matrices e.g. in
\cite{KDI-2000,Kwapien,BT}. Non-Hermitian Wishart ensembles also appeared
naturally as the described two-matrix problem in the study of the
Dirac operator spectrum of Quantum Chromodynamics (QCD) with chemical
potential $\mu$. Here the ensemble with complex ($\beta=2$)
\cite{Osborn}, quaternion real ($\beta=4$) \cite{A05}, and real
($\beta=1$) \cite{APSo,APSoII} matrix elements were first introduced
and solved for finite and infinite matrix dimensions 
by Osborn, the author and his coworkers, respectively.
In particular this includes the generalization of the Bessel kernels
in the microscopic origin scaling limit into the complex plane for all
three ensembles. 
We refer to \cite{Jac} for the most recent review on the topic of
random matrix applications to QCD. Because the solution of these three
ensembles can be expressed in terms of Laguerre polynomials in the
complex plane, and because these ensembles display chiral symmetry
(see e.g. \cite{Jac}) they are also called non-Hermitian (or complex)
Laguerre or chiral ensembles. 
The link to statistical applications of non-Hermitian Wishart
ensembles was reemphasized more recently in \cite{KS}. Here and
independently in \cite{Burda1} the spectral density generalizing the
Marchenko-Pastur distribution into the complex plane was computed. In
\cite{Burda1} the product of having also more than two rectangular
matrices was considered.

The two-matrix models constructed and solved for applications to QCD
are much more general, for two reasons. First, they depend on a
non-Hermiticity parameter (the chemical potential $\mu$) that allows
to smoothly deform the Hermitian Wishart (-Laguerre or chiral)
ensembles into the maximally 
non-Hermitian Wishart ensembles of two independent matrices described
above. In the Ginibre ensembles these deformations also exist and are
called elliptic, or Ginibre-Girko ensembles. The parameter $\mu$
allows 
to study deformations of the kernels known to be universal from the
Hermitian setting. Second, more source terms were added to these
parameter dependent two-matrix models, by inserting an arbitrary but
fixed number of characteristic polynomials (as additional determinants
of the Dirac operator) into the measure. 
All complex eigenvalue correlation functions were computed in this
more general setting in \cite{Osborn,AOSV} for $\beta=2$, in
\cite{A05,AB} for $\beta=4$ and in \cite{AKP,AKPW} for $\beta=1$. We
also mention that the $\beta=1$ symmetry class at maximal
non-Hermiticity appears in the superconducting phase of QCD with two
colors \cite{KWY}. 
In this short presentation we will focus on the first aspect, the
dependence on the non-Hermiticity parameter and recall the full
solution in terms of (skew) orthogonal polynomials in the complex
plane. The additional characteristic polynomials 
can then be easily implemented by
modifying the corresponding kernels (see e.g. in \cite{AOSV,AKP}).

Finally we would like to mention some related developments in
non-Hermitian random matrices. 
First of all non-Hermitian generalizations of Wishart ensembles were
first considered as one-matrix models \cite{Steph,HOV} where the
non-Hermiticity is provided by a constant matrix shift. These models
are much more difficult to be solved in general, and only the microscopic density
for $\beta=2$ was determined in \cite{KJreplica}. 

Instead of considering the complex eigenvalues of the product of two
matrices, in \cite{ForresterMays} those of the ratio of two matrices
were studied, which relate to a Cauchy distribution. When generalizing
from one to two Wishart matrices, one can also consider  the positive
Hermitian (real symmetric) combination $(DC)^\dag DC$, as was done in
for instance in \cite{Karol}, including more matrices. Again this
generalizes the Marchenko-Pastur density, this times for real
eigenvalues, finding generating functions relevant in combinatorics. 
More generally speaking, spectral properties of the product of
quadratic random matrices - Hermitian or non-Hermitian - have been
studied by several authors in the literature, and we refer to
\cite{ProdBook} as well as to the contribution \cite{Burda2} to these
proceedings and references therein. 
The three Ginibre ensembles, and the three non-Hermitian Wishart
ensembles reviewed here are not the only possible non-Hermitian random
matrices one can consider. For an ordering principle we refer to
\cite{BLeC,Magnea} which include these 6 classes out of 33
non-Hermitian ones. 
For recent reviews on non-Hermitian random matrices we refer to
\cite{BHJ,Zabrodin}.

This review is organized as follows. In Sect. \ref{Z} we recall the
definition of generalized non-Hermitian 
Wishart ensembles, discuss their relation to
standard Wishart ensembles and give their complex eigenvalues
representations. Sect. \ref{OPonC} gives a list of the three sets of
(skew) orthogonal Laguerre polynomials in the complex plane and their
kernels that allows to compute all complex (and real) eigenvalue
correlation functions. The large-$N$ limit is sketched in
Sect. \ref{largeN}, starting with the elliptic law for the Dirac
eigenvalues and then listing the three Bessel kernels in the
microscopic origin limit at weak and strong non-Hermiticity. A short
discussion on universality follows in Sect. \ref{open}.

\section{Complex eigenvalue representation of the partition function}
\label{Z}

We begin by recalling the definition of the standard Wishart ensembles
we wish to generalize. Its partition function is given by 
\be
Z_{N,\,\nu}^{(\beta)} \sim\int dC
\exp\left[-\frac{N}{2}\Tr (C^\dag C )
\right]\ ,
\label{Wishart}
\ee
where $C$ is a rectangular matrix of size $N\times(N+\nu)$ with real
($\beta=1$), complex ($\beta=2$) or quaternion real ($\beta=4$) matrix
element without further symmetry. 
When going to a real eigenvalue basis for these ensembles there are
two choices\footnote{We note in passing that on the level of matrix
  elements eq. (\ref{Wishart}) is identical to the Ginibre
  ensembles for $\nu=0$. The difference is that there the complex eigenvalues of
  the matrix $C$ are studied.}. Either we can consider the $N$
positive eigenvalues $\la_i$ of the positive Hermitian matrix $C^\dag
C $, which we will call Wishart eigenvalues (or Wishart picture). Or
we can consider the $2N+\nu$ eigenvalues $x_i$ 
of the Dirac matrix $\Dirac=\left(\begin{array}{cc}
\mbox{\bf 0}_N& C\\
C^\dag               & \mbox{\bf 0}_{N+\nu}
\end{array} \right)$ which we call Dirac eigenvalues (or Dirac
picture). When comparing the two characteristic equations, 
\be
\det[\la-C^\dag C]=\prod_{i=1}^N(\la-\la_i)\ , \ \ \mbox{vs.}\ \ 
\det[\la-\Dirac]=\la^\nu\prod_{i=1}^N(\la^2-x_i^2)\ ,
\label{characteristic}
\ee
it is clear that we have $\nu$ zero- and $2N$ non-zero Dirac
eigenvalues that come in $\pm$ pairs. The simple substitution
$\la_i=x_i^2$ will lead us from one picture to the other 
for the non-zero eigenvalues. When computing the standard Jacobian for
the diagonalization we obtain the partition function in terms of
eigenvalues as follows 
\bea
Z_{N,\,\nu}^{(\beta)} &\sim&\int_0^\infty \prod_{i=1}^N
d\la_i\la_i^{\frac{\beta}{2}(\nu+1)-1}e^{-N\la_i}\
|\Delta_N(\{\la_j\})|^\beta\nn\\ 
&\sim&\int_{-\infty}^\infty \prod_{i=1}^N
dx_i|x_i|^{{\beta}(\nu+1)-1}e^{-Nx_i^2}\ |\Delta_N(\{x_j^2\})|^\beta. 
\eea 
It is given in both the Wishart and Dirac picture, and
we have defined the Vandermonde determinant as
\be
\Delta_N(\{\la_j\})\equiv\prod_{N\geq k>l\geq 1}(\la_k-\la_l)
=\det_{1\leq k,l\leq N}\Big[\lambda_k^{l-1}\Big]\ .
\ee

Let us now turn to the non-Hermitian generalization. As sketched in
the introduction we consider a Gaussian two-matrix model and compute
the spectral properties of the non-Hermitian matrix $DC$: 
\be
{\cal Z}_{N,\,\nu}^{(\beta)}\sim \int dC dD
\exp\left[-\frac{N}{2}\Tr (C^\dag C +D^\dag D )
\right]\ .
\label{nHWishartmax}
\ee
Both $C$ and $D^\dag$ are rectangular matrices of size
$N\times(N+\nu)$ with real ($\beta=1$), complex ($\beta=2$) or
quaternion real ($\beta=4$) matrix elements without further symmetry. 
Instead of the $N$ complex eigenvalues $v_i$ of the product matrix
$DC$ we may again consider the spectrum of the non-Hermitian Dirac
matrix $\Dirac=\left(\begin{array}{cc} 
\mbox{\bf 0}_N& C\\
D               & \mbox{\bf 0}_{N+\nu}
\end{array} \right)$ instead. Here the ensemble ($\beta=1$) has a
special feature. The characteristic equation of the real asymmetric
matrix $DC$ remains real. Therefore the solutions of
$\det[\la-DC]=\prod_{i=1}^N(\la-v_i)$ are either real, or occur in
complex conjugate pairs. 
For the non-zero complex Dirac eigenvalues $z_i$ we thus get 3
different possibilities listed in Table \ref{tabCev}. 
\begin{table}
\begin{tabular}{l|l|l}
  & Wishart picture     & Dirac picture\\ \hline\hline
a)& $v_j>0$ real & $\pm z_j\in\mathbb{R}$, $z_j^2=v_j$\\ \hline

b)& $v_j<0$ real & $\pm z_j\in i\mathbb{R}$, $z_j^2=v_j$\\ \hline

c)& $v_j,\ v_j^*$ complex conjugate pair & $\pm z_j,\pm z_j^*\in
\mathbb{C}$, 2 pairs $z_j^{2\,(*)}=v_j^{(*)}$\\ \hline 

\end{tabular}
\caption{Non-zero complex eigenvalues for real matrix elements
  $\beta=1$.}\label{tabCev} 
\end{table}

In the following we will solve a more general non-Hermitian extension
of the Wishart ensembles than eq. (\ref{nHWishartmax}). We add a
non-Hermiticity parameter $\mu\in[0,1]$ that allows to interpolate
between the Wishart ensemble eq. (\ref{Wishart}) for $\mu=0$ and its
generalization eq. (\ref{nHWishartmax}) for $\mu=1$ which we call
maximally non-Hermitian. The partition function is defined as 
\bea
{\cal Z}_{N,\,\nu}^{(\beta)}(\mu;\{m_f\}) &\sim&
 \int dC  dD
\ \exp\left[-{N}\eta_+\Tr (C^\dag C + DD^\dag )\right]
\label{Z2MMCD}\\
&\times&
\exp\left[-{N}\eta_-\Tr (DC + C^\dag D^\dag)\right]
\prod_{f=1}^{N_f}
\det[{\mathbf 1}_{2N+\nu}m_f+\Dirac]\ ,\nn\\
\eta_\pm &\equiv& \frac{1\pm\mu^2}{4\mu^2}\ .
\label{etapm}
\eea
In the second line we have added a product of characteristic
polynomials to the weight function, these are called $N_f$ quark
flavors in the language of QCD. All complex (and real for $\beta=1$)
eigenvalue correlation functions as well as the partition functions
are known in the presence of these terms, see
\cite{Osborn,AOSV,A05,AB,APSo,AKP,AKPW,PhDMJP}. For simplicity we will
mainly restrict ourselves to the so-called quenched case $N_f=0$ in
the following, although $N_f\neq0$ is particularly interesting for the
QCD application as it may lead to a non-positive overall weight, the
so-called sign problem. For more discussion we refer to these
references, as well as to \cite{Jac}.

The fact that the ensembles eq. (\ref{Z2MMCD}) indeed extrapolate
between the maximally non-Hermitian and Hermitian Wishart ensembles
can be seen as follows. The change of variables 
\be
C \equiv \Phi\ + i\mu \Psi \ ,\ \ D \equiv \Phi^{\dagger} + i\mu
\Psi^{\dagger} \ , 
\label{CDdef}
\ee
leads to uncoupled Gaussian weights in the matrices $\Phi$ and $\Psi$
(see e.g. \cite{Osborn}). For $\mu=1$ this change of variables is
trivial and we are back to eq. (\ref{nHWishartmax}). Aspects of this
simpler model at $\mu=1$ were also treated in
\cite{APSo,KWY,KS,Burda1}. 
For $\mu=0$, $D=C^\dag$, and so $\Dirac$ becomes Hermitian and
independent of $\Psi$ which thus decouples, to give the standard
Wishart ensembles eq. (\ref{Wishart}) in terms of matrix $\Phi$. For
completeness and later use we also state the elliptic extension of the
Ginibre ensembles corresponding to eq. (\ref{Z2MMCD}) (at $N_f=0)$: 
\bea
{\cal Z}_{N,\,Gin}^{(\beta)}(\mu) \sim
 \int dJ
\ \exp\left[-Na\left\{\eta_+\Tr (JJ^\dag)
-\eta_-\Tr (J^2 + J^{\dag\,2})\right\}\right]\ ,
\label{Ginell}
\eea
where $J\neq J^\dag$ is an $N\times N$ non-Hermitian matrix. In
\cite{FKS98} we have 
$a=(1+v^2)$ and $\tau=(1-v^2)/(1+v^2)$.

Let us now turn to the complex eigenvalue representation of eq. (\ref{Z2MMCD}),
defining the joint probability distribution
\be
{\cal Z}_{N,\,\nu}^{(\beta)}(\mu)\equiv \int_{\mathbb{C}}  \prod_{j=1}^N d^2z_j
{\cal P}_{N,\,\nu}^{(\beta)}(\{z\})\ .
\label{Pjpdf}
\ee
Here we present the Wishart picture only.
For the complex and quaternion real case we obtain
\bea
{\cal Z}_{N,\,\nu}^{(\beta=2)}(\mu) &=&
\int_{\mathbb{C}}  \prod_{j=1}^N d^2z_j w_\nu^{(\beta=2)}(z_j)
|\Delta_{N}(\{z\})|^2\ ,
\label{Z2b2ev}\\
{\cal Z}_{N,\,\nu}^{(\beta=4)}(\mu) &=&
\int_{\mathbb{C}}  \prod_{j=1}^N d^2z_j w_\nu^{(\beta=4)}(z_j) (z_j-z_j^{*})\
\Delta_{2N}(\{z,z^{*}\})\ ,
\label{Z2b4ev}
\eea
where we have defined the weight function
\be
w_\nu^{(\beta)}(z)\equiv|z|^{\frac{\beta}{2}\nu}
K_{{\beta}\,\nu/2}\left(2N\eta_+|z|\right)
\exp\left[N\eta_-(z+z^{*})\right] .
\label{weight}
\ee
The appearance of the Bessel-$K$ function can be understood as
follows, just considering scalar variables. While the sum of two
random variables (the real and imaginary part) is again Gaussian 
- this corresponds to the Ginibre case - 
here we consider the product of two Gaussian random variables, which
is distributed with respect to Bessel-$K_0$ as one case easily
convince oneself. 
For more details of the derivation of the Jacobians we refer to
\cite{Osborn,A05}. Note that for $\beta=4$ the Jacobian is different
from $|\Delta_N(\{z\})|^4$ as one would expect for a standard
Dyson-gas \cite{Zabrodin}. For an interpretation in terms of charged
particles we refer to \cite{Forresterbook}. 
Very recently an alternative derivation of the Jacobian  for $\beta=2$
has been presented in \cite{KS}. 

In the third and most difficult ensemble $\beta=1$ with rectangular
real matrices one has in principle to sum over all possible
combinations of complex conjugate and real eigenvalues (and purely
imaginary ones in the Dirac picture) \cite{APSoII}. We only quote the
result  
for the partition function of \cite{AKP} where a factorized form was
shown, reading as follows: 
\be
{\cal Z}_{N=2n+\chi\,\nu}^{(\beta=1)}(\mu) =
\int\limits_{\mathbb{R}}dy^\chi\ h^\chi(y)
\prod_{k=1}^{2n}  \int\limits_{\mathbb{C}} d^{\,2}z_k
\prod\limits_{j=1}^{n}F_\nu(z_{2j-1},z_{2j})
\ \Delta_{\chi+2n}(y,\{z\})\ .
\label{Zgen}
\ee
The anti-symmetric weight function is defined as
\begin{eqnarray}
\label{Fdef}
F_\nu(z_{1},z_{2})
&\equiv&
{i}g_\nu(z_{1},z_{2})(\Theta(\im\,z_{1})-\Theta(\im\,z_{2}))\,
\delta^2(z_{2}-z_{1}^*)\\
&&+  \frac{1}{2}  h_\nu(z_{1})h_\nu(z_{2})\delta(\im\,z_{1})\delta(\im\,z_{2})
\sgn(\re\,z_{2}-\re\,z_{1})\, .\nn
\end{eqnarray}
It is given in terms of the following two weights for real and complex
eigenvalues respectively: 
\bea
\label{wch}
h_\nu(x) &\equiv&
2|x|^{\nu/2}K_{\nu/2}(2N\eta_+|x|)\exp[2N{\eta_-x}]
\ ,\\
g_\nu(z_1,z_2)&\equiv&
2|z_1z_2|^{\nu/2}\exp[2{N\eta_-(z_1+z_2)}]
\int_0^\infty \frac{dt}{t}e^{-4N^2\eta_+^2 t(z_1^2+z_2^2)-\frac{1}{4t}}
\nn\\
&&\times
K_{\nu/2}\left(8N^2\eta_+^2t z_1 z_2\right)
\erfc\left(2N\eta_+\sqrt{t}|z_2-z_1|\right).
\nn
\eea
These two weights are related by
\be
\lim_{\Im m \, z\to0}g_\nu(z,z^*)=h_\nu(\re \, z)^2\ .
\label{hg}
\ee
While both contain parts of the weight function eq. (\ref{weight})
with $\beta=1$ there, in particular the weight for complex eigenvalues
is more complicated here. 
In eq. (\ref{Zgen}) valid for both even and odd $N=2n+\chi$,
$n\in{\mathbb N}$ the notation indexed by $\chi$ means that for even
(odd)  $N$ with $\chi=0$ (1) the additional integration over the real
eigenvalue $y$ is absent (present). The fact that an absolute value
around the Vandermonde determinant is absent here is due to the
ordering enforced by the weight $F$ in terms of the sign- and
$\Theta$-functions. 
For details on the computation of the Jacobian we refer to \cite{APSoII,PhDMJP}.

\section{Correlation functions at finite $N$: (skew) orthogonal Laguerre
polynomials in the complex plane}
\label{OPonC}

We now turn to the computation of complex (and real) eigenvalue
correlations functions. The $k$-point density correlation functions
are defined as 
\be
R_N^{(\beta)}(z_1,\ldots,z_k)\ \equiv\ \frac{N!}{(N-k)!}
\frac{1}{{\cal Z}_{N,\,\nu}^{(\beta)}(\mu)}
\int_{\mathbb C}\prod_{j=k+1}^N  d^2z_j  {\cal P}_{N,\,\nu}^{(\beta)}(\{z\}) \ .
\label{defRk}
\ee
The map to the density correlations of Dirac eigenvalues $(\Dirac)$ is
then given by 
\be
R_{N}^{(\beta,\,\Dirac)}(z_1,\ldots,z_k) \ =\
2^{2k}\prod_{j=1}^k|z_j|^2\ R_{N}^{(\beta)}(z_1^2,\ldots,z_k^2)\ . 
\ee
For all three ensembles these can be solved in terms of a kernel
defined in term of Laguerre polynomials in the complex plane as we
will show now. 
Analogous results hold for the Ginibre ensembles with $\beta=2,4$ and
1 in terms of Hermite polynomials in the complex plane as shown in 
\cite{FKS98,EK,Forrester07}, respectively. 
The case $\beta=1$ is again special as the $k$-point function will
consist of a sum of all 
possibilities of real and complex conjugate eigenvalue pairs of a
total number $k$ (see e.g. \cite{APSoII}). 
Other correlation functions can be defined and computed in these
ensembles as well such
as gap probabilities, and we refer to \cite{APS} for more details. 

For $\beta=2$ the $k$-point correlation functions can be solved in
terms of the kernel 
of orthogonal polynomials, in complete analogy to the Hermitian case:
\be
R_N^{(\beta=2)}(z_1,\ldots,z_k)\ = \prod_{l=1}^k w_\nu^{(\beta=2)}(z_l)\
\det_{1\leq i,j\leq k}\left[ {\cal K}_N^{(\beta=2)}(z_i,z_j)\right]\ .
\label{Rkbeta2}
\ee
For the weight eq. (\ref{weight}) it is given through the orthonormalized
Laguerre polynomials in the complex plane \cite{Osborn,AOSV} 
\be
{\cal K}_N^{(\beta=2)}(z,u)
=
\sum_{k=0}^{N-1}
\left(\frac{\eta_-}{\eta_+}\right)^{2k}
\frac{N^{\nu+2}k!}{\pi(1+\mu^2)^{\nu}(k+\nu)!}
L_k^{\nu}\left(\frac{Nz}{4\mu^2\eta_-}\right)
L_k^{\nu}\left(\frac{Nu^*}{4\mu^2\eta_-}\right).
\label{Kbeta2}
\ee
The Laguerre polynomials satisfy the following general orthogonality relation
\be
\int_{\mathbb{C}}dz^2
|z|^{\nu} K_\nu\left( a|z|\right)\exp\left[\frac{b}{2}(z+z^*)\right]
L_j^{\nu}\left(cz\right)
L_k^{\nu}\left(cz^*\right)
=h_j^{(2)}\delta_{jk}\ ,
\ee
for $a>b>0$ with $c\equiv (a^2-b^2)/(2b)$. The squared norms read
\be\label{norm}
h_j^{(2)}=\frac{\pi(j+\nu)!}{a\ j!}\left( \frac{a}{b}\right)^{2j}
\left(\frac{2a}{a^2-b^2}\right)^{\nu+1} .
\ee
A short proof for this relation stated in \cite{Osborn} can be
found in Proposition 1 in \cite{ABe} (see also appendix A of
\cite{A05} for an earlier proof). 
The simplest example for a correlation function  is the spectral density
\be
R_N^{(\beta=2)}(z)\ = \ w_\nu^{(\beta=2)}(z)\
{\cal K}_N^{(\beta=2)}(z,z^*)\ ,
\label{R1b2}
\ee
with the kernel from eq. (\ref{Kbeta2}), see Fig. \ref{rhoCbeta2}. At
maximal non-Hermiticity $\mu=1$ it reduces to 
an incomplete Bessel-$I$ function times the Bessel-$K$ from the weight
\cite{AOSV} 
\be
R_N^{(\beta=2)}(z_1)\Big|_{\mu=1}\ = \ K_\nu(N|z|)\sum_{l=0}^{N-1}
\frac{N^2}{\pi(l+\nu)!l!}\left(\frac{N|z|}{2}\right)^{2l+\nu}\ . 
\label{R1b2max}
\ee
For more details about correlations functions including $N_f$
characteristic polynomials we refer to \cite{Osborn,AOSV}. 

For $\beta=4$ the $k$-point correlation functions are given by
\be
R_N^{(\beta=4)}(z_1,\ldots,z_k)\ = \prod_{l=1}^k
w_\nu^{(\beta=4)}(z_l)(z_l^*-z_l)\ 
\mbox{Pf}_{i,j=1,\ldots,2k}
\left[{\cal K}_N^{(\beta=4)}(u_i,u_j)\right] \ ,
\label{detformula}
\ee
where the arguments of the antisymmetric kernel 
\be
{\cal K}_N^{(\beta=4)}(z_1,z_2^\ast)\ \equiv\
\sum_{k=0}^{N-1} \frac{1}{h_k^{(4)}}
\left(q_{2k+1}^{(\beta=4)}(z_1)q_{2k}^{(\beta=4)}(z_2^\ast)- 
q_{2k+1}^{(\beta=4)}(z_2^\ast)q_{2k}^{(\beta=4)}(z_1)\right)  \ ,
\label{prekernel}
\ee
run through the set of $2k$ variables
$\{u_j\}=\{z_1,z_1^*,\dots,z_k,z_k^*\}$, and the Pfaffian of an
antisymmetric matric of size $2k$ is given by the square root of the
determinant of the matrix. 
The polynomials inside the kernel are given by \cite{A05}
(with $q_{2k+1}^{(\beta=4)}(z)$ determined up to a constant times
$q_{2k}^{(\beta=4)}(z)$) 
\bea
q_{2k+1}^{(\beta=4)}(z)&=& -(2k+1)!\left(\frac{4\mu^2\eta_-}{N}\right)^{2k+1}
L_{2k+1}^{2\nu}\left(\frac{Nz}{4\mu^2\eta_-}\right)\ ,
\label{qodd}\\
q_{2k}^{(\beta=4)}(z)&=& \left(\frac{8\mu^2\eta_+}{N}\right)^{2k}
\sum_{j=0}^k \left(\frac{\eta_-}{\eta_+}\right)^{2j}
\frac{k!\ (k+\nu)!(2j)!}{2^{2j}j!\ (j+\nu)!}
L_{2j}^{2\nu}\left(\frac{Nz}{4\mu^2\eta_-}\right),\nn
\label{qeven}
\eea
with squared norms
\be
h_k^{(4)}
\ =\ 8\pi\mu^4 (2k+1)!\ (2k+2\nu+1)!
\frac{{(1+\mu^2)^{4k+2\nu}}}{N^{4k+2\nu+4}}
\ .
\label{rkK}
\ee
They enjoy the following skew orthogonality conditions
\bea
\langle q_{2k+1}^{(4)}|q_{2l}^{(4)}\rangle_{4} &=& -\langle
q_{2l}^{(4)}|q_{2k+1}^{(4)}\rangle_{4} 
\ =\ h_k^{(4)}\ \delta_{kl}\nn\ ,\\
\langle q_{2k+1}^{(4)}|q_{2l+1}^{(4)}\rangle_{4} &=& \, \ \ \langle
q_{2l}^{(4)}|q_{2k}^{(4)}\rangle_{4} \ 
\ \ \ =\ 0\ ,
\label{skewdef4}
\eea
with respect to the following skew-product
\be
\langle h|g\rangle_{\beta=4}\ \equiv\ \int d^2z\ w_\nu^{(\beta=4)}(z)\
(z^{\ast}-z)[h(z)g(z)^\ast-h(z)^\ast g(z)]\ .
\label{scalar}
\ee
Hence they are called skew orthogonal Laguerre polynomials in the complex plane.
Again we give the simplest example, the spectral density plotted in
Fig. \ref{rhoCbeta4} 
\be
R_N^{(\beta=4)}(z)\ = w_\nu^{(\beta=4)}(z)(z^*-z){\cal K}_N^{(\beta=4)}(z,z^*)
\ ,
\label{R1b4}
\ee
obtained by inserting the kernel eq. (\ref{prekernel}). At maximal
non-Hermiticity it reduces to 
\cite{A05}
\bea
R_N^{(\beta=4)}(z)\Big|_{\mu=1}\
&=&|z|^{2\nu}K_{2\nu}(N|z|)(z^*-z)\frac{N^{2\nu+2}}{\pi2^{2\nu+3}} 
\label{R1b4max}\\
&&\times\sum_{k=0}^{N-1}\sum_{j=0}^k
\frac{k!(k+\nu)!(z^{2k+1}z^{*\,2j}
\ -\ z^{2j}z^{*\,2k+1})}{(2k+2\nu+1)!(2k+1)!2^{4j}j!(j+\nu)!}
.\nn
\eea

Because of the prefactor the density vanishes identically along the
real line, or after mapping to Dirac eigenvalues along the real and
imaginary axis. 
For results for the correlation functions including $N_f$ pairwise
degenerate and nondegenerate flavors we refer to \cite{A05} and
\cite{AB}, respectively. 

For $\beta=1$ we will only give the results for
the $k$-point correlation functions for even $N=2n$. They
are give by \cite{APSoII}
\bea
R_N^{(\beta=1)}(z_1,\ldots,z_k)= \Pf\left[
\begin{array}{cc}
{\cal K}_N(z_i,z_j)& -G_N(z_i,z_j)\\
G_N(z_j,z_i) & -W_N(z_i,z_j)
\end{array}
\right]_{1 \leq i,j \leq k}.
\label{RnPf}
\eea
In addition to the antisymmetric weight eq. (\ref{Fdef}) we have
introduced the following two functions of two complex variables 
\bea
G_N(z_1,z_2)&=&-\int_{\mathbb{C}} d^2z \, {\cal
  K}_N^{(\beta=1)}(z_1,z){F_\nu}(z,z_2)\ , 
\label{Gdef}\\
W_N(z_1,z_2)&=& -{F_\nu}(z_1,z_2)+\int_{\mathbb{C}}  d^2z
\int_{\mathbb{C}} d^2z'{F_\nu}(z_1,z)
{\cal K}_N^{(\beta=1)}(z,z'){F_\nu}(z',z_2)\ .
\nn
\eea
These are given in terms of the integrals of the antisymmetric kernel
\cite{APSo}
\bea
{\cal K}_N^{(\beta=1)}(z_1,z_2)&=&\frac{\eta_-N^\nu}{4\pi(8\mu^2\eta_+)^{\nu+1}}
\sum_{j=0}^{N-2}\left(\frac{\eta_-}{\eta_+}\right)^{2j}\!
\frac{(j+1)!}{(j+\nu)!}
\label{Kdefb1}\\
&\times&\left\{
L_{j+1}^\nu\Big( \frac{Nz_2}{8\mu^2\eta_-}\Big)
L_{j}^\nu\Big( \frac{Nz_1}{8\mu^2\eta_-}\Big)
-
(z_1\leftrightarrow z_2)
\right\}.
\nn
\eea
For odd $N$ the corresponding results can be obtained for example by
the limiting procedure proposed in \cite{ForresterMays}, or by
following \cite{Sommers2008}.

The above kernel can be expressed in terms of the following skew
orthogonal polynomials \cite{AKP}
\bea
q_{2k}^{(\beta=1)}(z)    & = & +C_{2k}^{\nu}(z),  \label{Lfinal} \\
q_{2k+1}^{(\beta=1)}(z)  & = & -C_{2k+1}^{\nu}(z) +
N^{-2}(1+\mu^2)^2\,(2k)(2k+\nu)C_{2k-1}^{\nu}(z) + c'C_{2k}^{\nu}(z)\ ,
\nn
\eea
giving the skew-orthogonal Laguerre polynomials up to an arbitrary
constant $c'$.
Here we have defined
\bea
C_k^{\nu}(z) & \equiv & \left(\frac{8\mu^2\eta_-}{N}\right)^{k}\,k!\,
L_k^{\nu}\left( \frac{Nz}{8\mu^2\eta_-} \right)\ .
\eea
The corresponding skew-product is defined as
\begin{equation}\label{skewdef1}
 \langle f|g\rangle_{\beta=1}=-
\langle g|f\rangle_{\beta=1}\equiv
\int d^{\,2}z_1\,d^{\,2}z_2\,
F_\nu(z_1,z_2)\det\left[\begin{array}{cc}
    f(z_1) & g(z_1) \\ f(z_2) & g(z_2) \end{array}\right],
\end{equation}
for two functions $f(z)$ and $g(z)$ that are integrable with
respect to the weight functions contained in $F_\nu(z_1,z_2)$ in 
eq. (\ref{Fdef}).
Our skew-orthogonal polynomials defined above then satisfy
\bea
\langle q_{2k}^{(1)}|q_{2l+1}^{(1)}\rangle_1&=&\ -\ \ \langle
q_{2l+1}^{(1)}|q_{2k+1}^{(1)}\rangle_1=\ h_k^{(1)}\delta_{kl}\ ,\nn\\ 
\langle q_{2k}^{(1)}|q_{2l}^{(1)}\rangle_1&=&\langle
q_{2k+1}^{(1)}|q_{2l+1}^{(1)}\rangle_1 
\ =\ 0 \ \forall k,l\geq0\ ,
\label{qdef}
\eea
where the $h_k^{(1)}>0$ are their positive (squared skew) norms
\be
h_k^{(1)}\equiv 8\pi(4\mu^2)
(2k)!\,(2k+\nu)!\,(8\mu^2\eta_+/N)^{4k+\nu+1}\ .
\ee
The kernel is then reading
\be
\label{kernelqq}
{\cal K}_{N=2n}^{(\beta=1)}(z_1,z_2) = \sum_{k=0}^{n-1} \,
\frac{1}{h_k^{(1)}} \, \big( 
q_{2k+1}^{(1)}(z_1)q_{2k}^{(1)}(z_2) -
q_{2k+1}^{(1)}(z_2)q_{2k}^{(1)}(z_1) \big), 
\ee
leading to eq. (\ref{Kdefb1}). 

As already mentioned, here the $k$-point function in eq. (\ref{defRk}) is
the sum of all possible combinations of complex and real eigenvalues,
starting from only complex to only real eigenvalues. For example for
the spectral density plotted in Fig. \ref{rhoCRbeta1} we have
\cite{APSoII} 
\be
R_{N}^{(1)}(z)= \int_{\mathbb{C}}  d^2z^\prime \, {\cal
  K}_N^{\beta=1}(z^\prime,z){F}(z,z^\prime) 
\equiv R_{N,\mathbb{C}}^{(1)}(z)+\delta(y)R_{N,\mathbb{R}}^{(1)}(x)\ .
\label{density}
\ee
Here we explicitly give the result obtained after inserting the kernel
(\ref{Kdefb1}) and integrating with respect to the two different
weights eq. (\ref{wch}). Denoting 
$z=x+iy$ we get in terms of the kernel from eq. (\ref{Kdefb1})
\bea
R_{N,\,\mathbb{C}}^{(1)}(z)&=& -2i(N|z|)^{\nu}e^{2N\eta_- x}
\sgn(y){\cal K}_N^{(1)}(z,z^*)2\int_0^\infty
\frac{dt}{t}e^{-2N^2\eta_+^2t(x^2-y^2)-\frac{1}{4t}} 
\nn\\
&&\times
K_{\frac{\nu}{2}}\Big(2N^2\eta_+^2t(x^2+y^2)\Big)
\erfc\left(2N\eta_+\sqrt{t}|y|\right) ,
\label{RcomplN}\\
R_{N,\,\mathbb{R}}^{(1)}(x)&=&4N^\nu
\int_{-\infty}^\infty dx' \sgn(x-x')\,
|xx'|^{\frac{\nu}{2}} e^{N\eta_- (x+x')}
K_{\frac{\nu}{2}}(N\eta_+|x|)
\nn\\
&&\times K_{\frac{\nu}{2}}(N\eta_+|x'|){\cal K}_N^{(1)}(x,x').
\label{RrealN}
\eea
At maximal non-Hermiticity the kernel 
eq. (\ref{Kdefb1}) considerably simplifies \cite{APSo}
\be
{\cal K }_N^{(1)} (z
,z^*)\Big|_{\mu=1}=\frac{1}{4\pi}\left(\frac{N}{4}\right)^{\nu+1} 
(z^*-z)
\sum_{l=0}^{N-2}\frac{1}{l!\,(l+\nu)!}
\left(\frac{N|z|}{2}\right)^{2l}\!.
\label{Kchmax}
\ee
Finally in the map to the Dirac picture one has to distinguish between
substituting complex or real eigenvalues. For the densities we have 
\bea\label{D-Wtrafob1}
R^{(1\, \Dirac)}_{N,\,\mathbb{C}}(z) & = & 4|z|^2
R^{(1)}_{N,\,\mathbb{C}}(z^2), \\ 
R^{(1,\, \Dirac)}_{N,\,\mathbb{R}}(x) & = & 2|x|
R^{(1)}_{N,\,\mathbb{R}}(x^2)\ , 
\label{D-Wtrafobreal}
\eea
and analogous relations hold for higher $k$-point functions.

\section{The large-$N$ limit: 3 classes of complex Bessel kernels}
\label{largeN}

\begin{figure}[h]
  \unitlength1.0cm
\centerline{
\epsfig{file=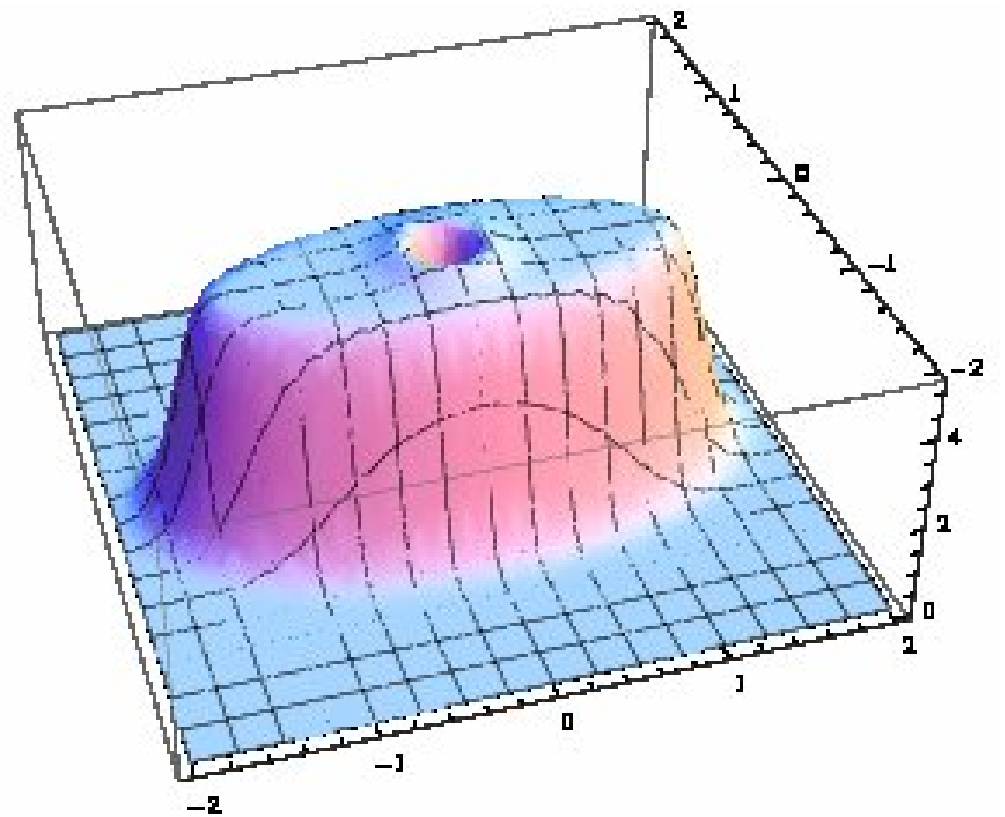,clip=,width=5.8cm}
\epsfig{file=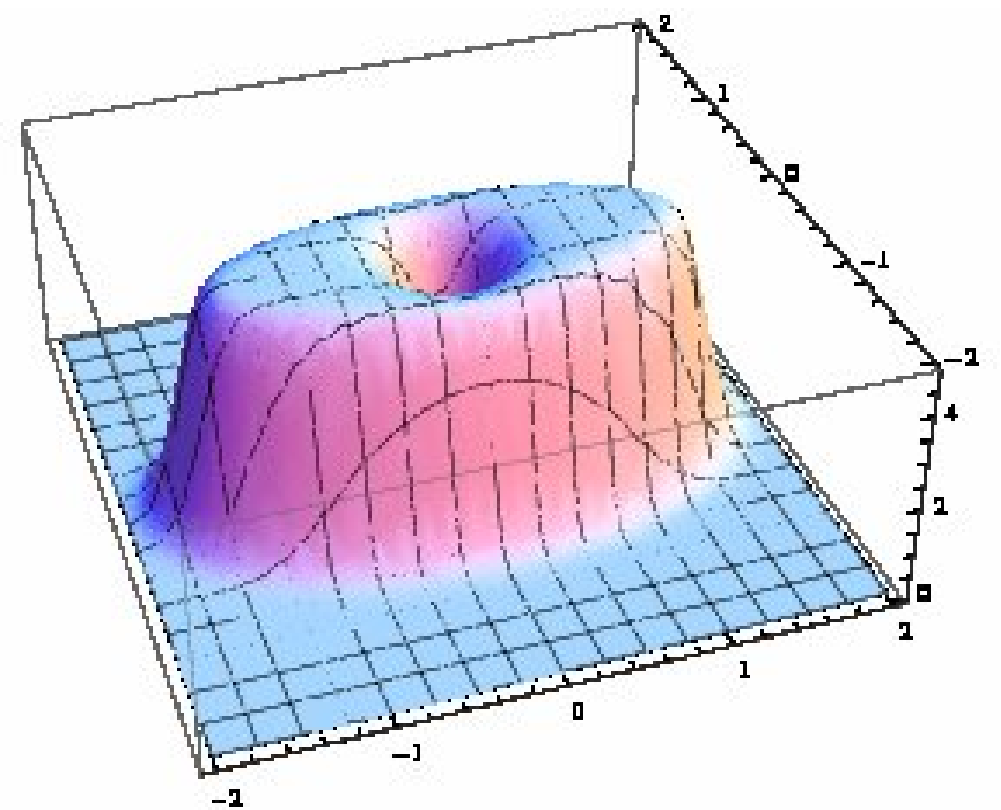,clip=,width=5.8cm}
}
  \caption{\label{rhoCbeta2}
The Dirac spectral density $R_N^{(\beta=2,\,\Dirac)}(z)$ for $\beta=2$ from
eq. (\ref{R1b2}) 
for $N=10$, $\mu=0.7$ and $\nu=0$ (left), and $\nu=1$ (right). The
extra repulsion from the origin through the zero-eigenvalue
at $\nu=1$ is clearly visible. In the large-$N$ limit the detailed structure
at the origin will only be visible on a microscopic scale.} 
\end{figure}

\begin{figure}[h]
  \unitlength1.0cm
\centerline{
\epsfig{file=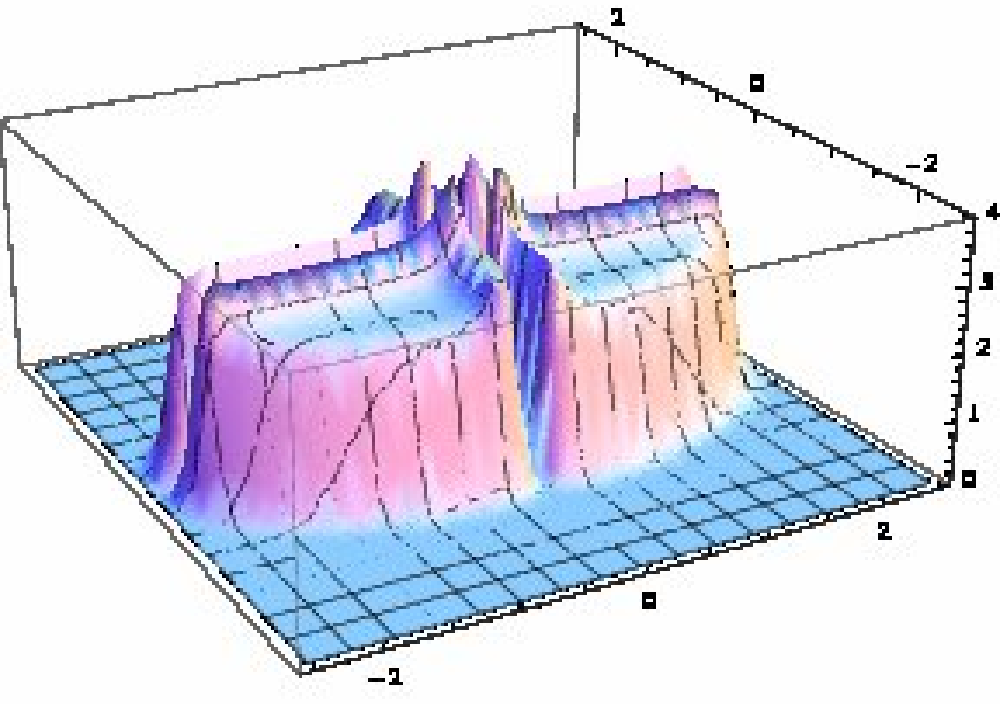,clip=,width=6.2cm}
\epsfig{file=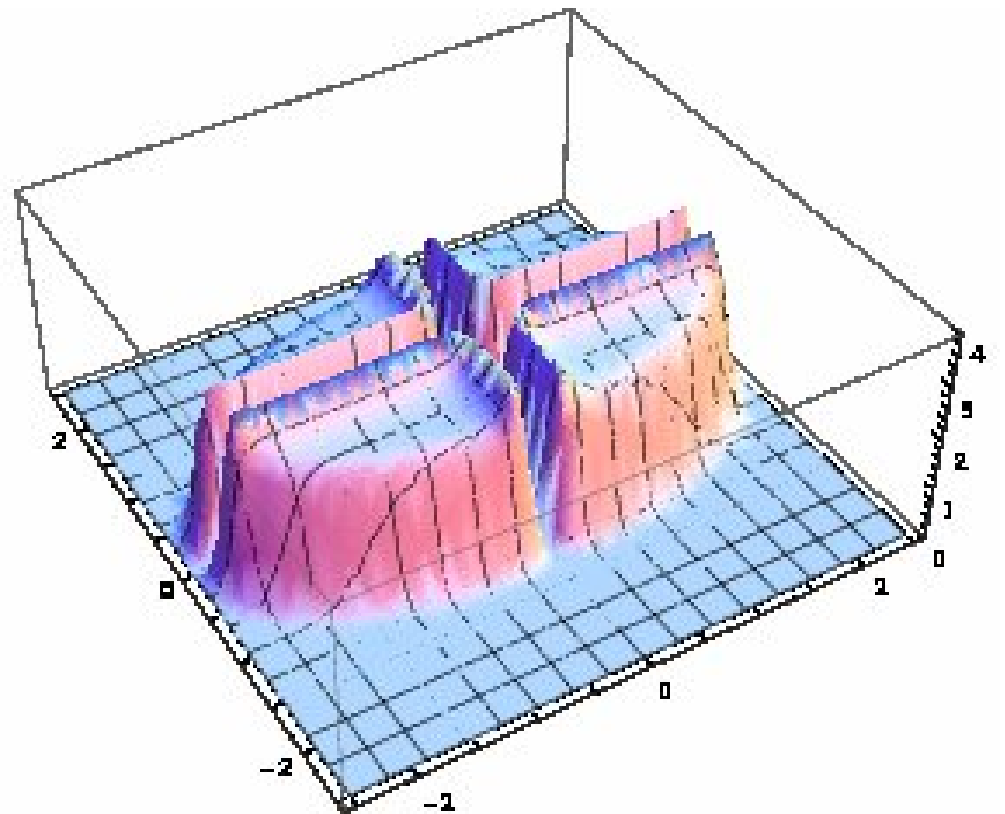,clip=,width=6.2cm}
}
  \caption{\label{rhoCbeta4}
The Dirac spectral density $R_N^{(\beta=4,\,\Dirac)}(z)$ for $\beta=4$ from
eq. (\ref{R1b4}) 
for $N=20$, $\mu=0.7$ and $\nu=0$ (left) and $\nu=1$ (right). In order
to see a smoother plateau we have increased $N$ here.  The extra
repulsion from the origin for $\nu=1$ is less pronounced here compared
to $\beta=2$.} 
\end{figure}

\begin{figure}[h]
  \unitlength1.0cm
\centerline{
\epsfig{file=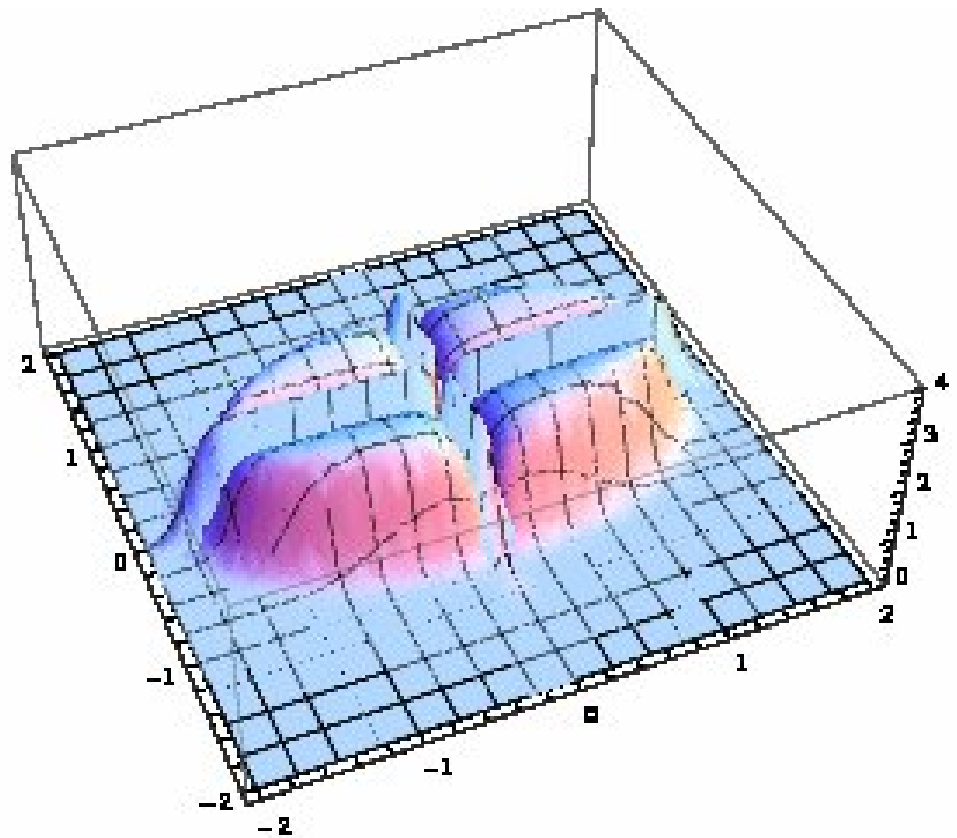,clip=,width=6.2cm}
\epsfig{file=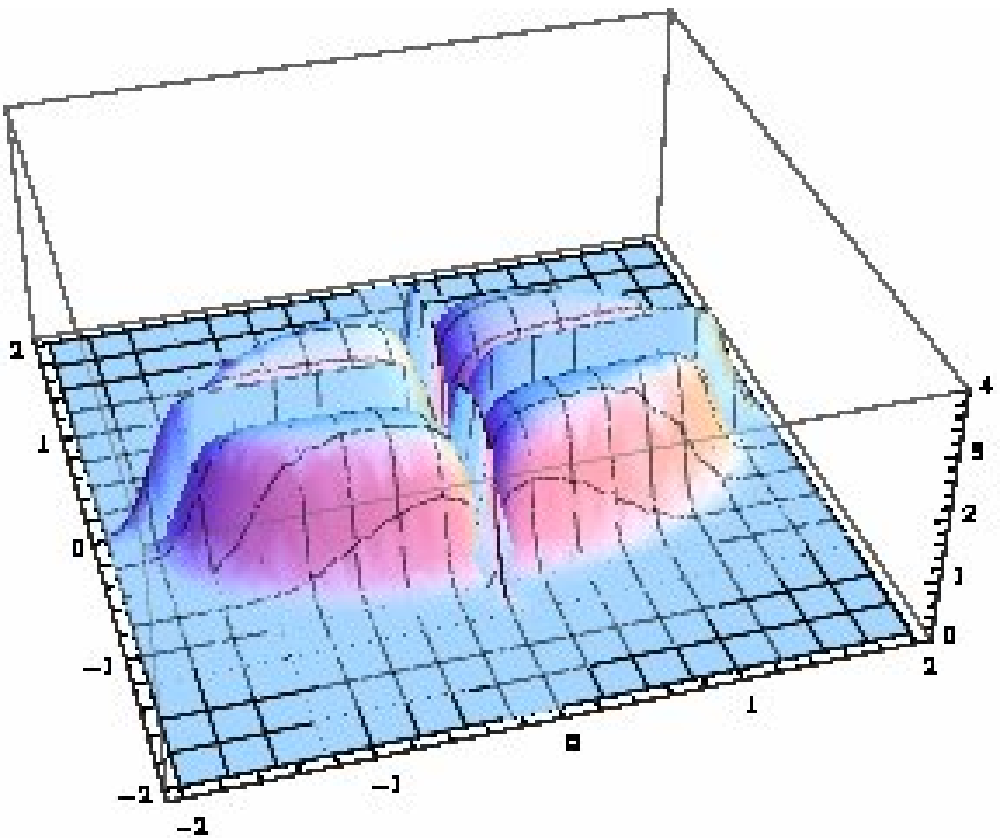,clip=,width=6.2cm}
}
  \caption{\label{rhoCRbeta1}
For $\beta=1$  we show two Dirac spectral densities: $R_{N,\mathbb
  C}^{(\beta=1,\,\Dirac)}(z)$
for the complex eigenvalues, and 
$R_{N,(i)\mathbb R}^{(\beta=1,\,\Dirac)}(z=x)$ $((iy))$
for the real (or purely
imaginary) eigenvalues from eqs. (\ref{RcomplN}) and (\ref{RrealN}). The
parameter values are the same as for $\beta=4$ in Fig. \ref{rhoCbeta4}. 
It is clearly visible that the density of complex eigenvalues is
repelled from the real and imaginary axis as for $\beta=4$, with an
apparently different profile though. 
}
\end{figure}

In this section we will review aspects of the large-$N$ limit for
non-Hermitian Wishart 
random matrices. Because the eigenvalues live in the complex plane we
have more possibilities in taking the large-$N$ limit, depending on
the location in the spectrum. 
In order to illustrate where such regions are located in the complex
plane we first display the so-called macroscopic (or global) limit of
the spectral density. As one can see from plotting the density at
finite-$N$ for all three $\beta=1,2,4$ in the Dirac picture in
Figs. \ref{rhoCbeta2} to \ref{rhoCRbeta1}, the global spectral density
follows the elliptic law \cite{Girko} as their Ginibre
counterparts. On top of being constant on an ellipse, various inner
and outer edges exist when zooming into particular regions. This is in
contrast to the real density of Wishart matrices where one can only
distinguish the (soft) edge, the bulk and the origin (or hard edge)
region. 

In addition to the location of the spectrum two different large-$N$
limits have to be distinguished for complex eigenvalues, that of
strong and weak non-Hermiticity \cite{FKS}. 
While in the former the eigenvalues fill a two-dimensional support, in
the latter they only locally extend into the complex plane. The weakly
non-Hermitian limit turns out to be a one-parameter deformation both
of the Hermitian and the strongly non-Hermitian limit. This makes it
an ideal object to study the question of universality. 

In the following we will focus on the microscopic limit at the origin,
as this is a special  feature of Wishart ensembles that their
corresponding Ginibre counterparts cannot offer.

We begin once more by recalling the results for the macroscopic
spectral density in Hermitian Wishart ensembles. Here two cases have
to be distinguished. Taking the large-$N$ limit with $\nu={\cal
  O}(N)$, such that $c\equiv N/(N+\nu)<1$ one obtains the
Marchenko-Pastur distribution \cite{MP} for the positive Wishart
eigenvalues for all three values of $\beta$ 
\be
\rho^{(\beta,M\!P)}(\la)=\frac{1}{2\pi c
  \la}\sqrt{(\la-cX_-)(cX_+-\la)}\ , \ \
X_\pm\equiv\sqrt{c^{-1/2}\pm1} 
\label{MP}
\ee
on $\la\in[cX_-,cX_+]$ and zero outside. For asymptotically square
matrices with $\nu={\cal O}(1)$ at large-$N$ (leading to $c=1$) it reduces to 
\be
\rho^{(\beta,M\!P)}(\la)=\frac{1}{2\pi}\sqrt{\frac{4}{\la}-1}
\ \Rightarrow\ \
\rho^{(\beta,\Dirac)}(\la)=2|\la|\rho^{(\beta,M\!P)}(\la^2) =
\frac{1}{\pi}\sqrt{4-\la^2}, 
\ee
which is just the semi-circle after mapping to the Dirac
picture. While in the Wishart picture we have a square root
singularity at the origin, the density in the Dirac picture is regular
at the origin. 

The same phenomenon happens in the complex plane. The semi-circle
density becomes replaced by the elliptic law, a constant density of an
ellipse \cite{Girko}, which we state for
the elliptic Ginibre ensemble eq. (\ref{Ginell}) 
(see e.g. \cite{FKS98}) in variables
$z=x+iy$ 
\be
\rho^{(Gin)}(z)=\frac{(1+v^2)^2}{4\pi v^2}\ \
\mbox{for}\ \  \frac{x^2}{4}(1+v^2)^2+\frac{y^2}{4v^2}(1+v^2)^2\leq 1,
\ee
and zero outside.
Away from the origin the mean spectral density of our three
non-Hermitian Wishart ensembles 
is constant on an ellipse in the Dirac picture, just as its non-chiral
counter parts. 
This can be seen in the figures below Fig. \ref{rhoCbeta2} -
\ref{rhoCRbeta1}, where we plot the corresponding spectral densities
after mapping to the Dirac picture according to eq. (\ref{D-Wtrafob1})
(and eq. (\ref{D-Wtrafobreal}) for $\beta=1$).

When mapping back to the Wishart picture using eq. (\ref{D-Wtrafob1})
we obtain a linear decay inside the support, as was observed in
\cite{KS,Burda1}: 
\be
\rho^{(\beta,\,\Dirac)}(z)\sim const.\ \ \Rightarrow\ \ \rho^{(\beta)}(z)
\sim \frac{const.}{|z|}\ .
\label{Well}
\ee
Likewise the behavior at the origin, that was found to be constant for
$\nu>0$ and logarithmically divergent $\sim\log|z|$ for $\nu=0$ in the Wishart
picture \cite{KS}, gets bent to zero in the Dirac picture using the map
eq. (\ref{D-Wtrafob1}). This can bee seen in the figures for all three
$\beta$ and the values of $\nu=0,1$. 

In addition, in \cite{KS,Burda1} the limit $\nu={\cal O}(N)$, such that
$q\equiv\nu/N>0$ is finite, was performed for $\beta=2$ at $\mu=1$, leading to 
the density 
\be
\rho^{(\beta=2)}(z)=\frac{1}{4\pi\sqrt{|z|^2+q^2}}\ ,
\ee
on the support, which we give in the Wishart picture here. This is the
generalization of the Marchenko-Pastur distribution eq. (\ref{MP})
into the complex plane. 

Let us now turn to the microscopic large-$N$ limit. As can be seen
from the figures several regions could be investigated by zooming into
the outer edges or inner edges along the axes. However, in the
following we will restrict ourselves to the microscopic origin limit,
as this limit is particular to the Wishart ensembles. 

We begin with the microscopic origin limit at strong non-Hermiticity
$(S)$. Because strong and maximal non-Hermiticity at $\mu=1$ are
related by a simple rescaling of the complex eigenvalues, $z\to
2\eta_+z$ (see e.g. \cite{FKS98,APSoII}), we restrict ourselves only to
the latter. 
This limit is defined such that the non-Hermiticity parameter $\mu$ is
not scaled with $N$, only the complex eigenvalues get rescaled: 
\be
\rho^{(\beta,\,\Dirac)}_S(\xi)\ \equiv\ \lim_{N\to\infty}\frac{1}{N}
R_N^{(\beta,\,\Dirac)}\left(z=\xi/\sqrt{N}\right)\ ,\ \
\lim_{N\to\infty} \sqrt{N}z=\xi \ \ \mbox{fixed}, 
\label{strdef}
\ee
and all higher order correlation functions are rescaled accordingly.

For $\beta=2$ we obtain easily from eq. (\ref{R1b2max}) \cite{AOSV}
\be
\rho^{(\beta,\,\Dirac)}_S(\xi)
\ =\ \frac{2}{\pi}|\xi|^2
K_\nu(|\xi|^2)I_\nu(|\xi|^2)\stackrel{|\xi|\to\infty}{\longrightarrow} 
\frac{1}{\pi}\ .
\label{rhob2str}
\ee
It only depends on the rescaled modulus $|\xi|$ and is thus rotationally
invariant. We also give the asymptotic value of the microscopic
density that matches the constant value of the macroscopic density in
the Dirac picture. 

For $\beta=4$ we obtain from eq. (\ref{R1b4max}) 
after a non-trivial calculation \cite{A05}
\bea
\rho^{(4,\,\Dirac)}_S(\xi)
&=&
\frac{|\xi|^2}{4\pi} (\xi^{\ast\,2}-\xi^2)
K_{2\nu}\left({|\xi|^2}\right)
\label{rhob4str}\\
&&\times
\int_0^1du\frac{I_{2\nu}(u|\xi|^2)}{\sqrt{1-u^2}}
\sinh\left(\frac{\sqrt{1-u^2}}{2}(\xi^2-\xi^{\ast\,2})\right) .
\nn
\eea
Compared to $\beta=2$ the density is no longer rotationally
invariant, reflecting the repulsion of the complex eigenvalues from
real and imaginary axes. 

For $\beta=1$ we obtain two densities from eqs. (\ref{RcomplN}) and
(\ref{RrealN}), for the complex eigenvalues and the eigenvalues on the
real and imaginary axis \cite{APSoII}: 
\bea
\rho_{\mathbb{C},S}^{(1,\,\Dirac)}(\xi) & = & \sgn(\Im m(\xi^2))
(-i)(\xi^2-\xi^{*\,2})\frac{8}{\pi}
|\xi|^2I_{\nu}\left( 2|\xi|^2 \right)
\nn\\
 && \times \int_{0}^{\infty} \frac{dt}{t} \,
e^{-(\xi^4+\xi^{*\,4}){t} -
\frac{1}{4t}} \,
K_{\frac{\nu}{2}}\left(2{|\xi|^4 t}\right)  \erfc\left(
2\sqrt{t}\,|\Im m(\xi^2)|\right)\!,\ \ \ \ \ \ 
\label{rhob1Cstr}\\
\rho_{(i)\mathbb{R}\,S}^{(1,\,\Dirac)}(\xi) & = & \frac{1}{\pi}|\xi|
 \, K_{\frac{\nu}{2}}(|\xi|^2) \left(
\int_{0}^{\infty}dx'\,|\xi^2 -x'|\
K_{\frac{\nu}{2}}(|x'|)\
I_{\nu}(2\xi \sqrt{x'})\right.
\nn\\
&&
\left.
+\int_{-\infty}^{0}dx'\,|\xi^2 -x'|\
K_{\frac{\nu}{2}}(|x'|)\
J_{\nu}( 2\xi\sqrt{|x'|})
\right),
\label{rhob1Rstr}
\eea
where in the second equation $\xi\in{\mathbb R}$ (or $\xi\in i{\mathbb
  R}$ which is identical here). Once again the density in the complex
plane is not rotationally invariant, and vanishes along the real and
imaginary axis. 

We now turn to the microscopic origin limit as weak non-Hermiticity
$(W)$ first introduced for the Ginibre ensembles eq. (\ref{Ginell})
in \cite{FKS}. It is defined by both rescaling $\mu$
with $N$, as well as the complex eigenvalues (with a different power
compared to the strong limit above): 
\be
\rho^{(\beta,\,\Dirac)}_W(\xi)\ \equiv\ \lim_{N\to\infty}\frac{1}{N^2}
R_N^{(\beta,\,\Dirac)}\left(z=\xi/{N}\right)\ ,
\ee
with
\be
\lim_{N\to\infty} 2{N^2}\mu^2=\alpha^2\ 
\ee
fixed. All higher order correlation functions are rescaled accordingly.

For $\beta=2$ we get \cite{Osborn}
\bea
\rho_W^{(2,\,\Dirac)}(\xi)&=&
\frac{1}{2\pi\alpha^2}|\xi|^2 K_\nu\left(\frac{|\xi|^2}{4\alpha^2}\right)
e^{\frac{\xi^2+\xi^{*\,2}}{8\alpha^2}}
\int_0^1dt\, t\, e^{-2\alpha^2t^2}J_\nu(t\xi)J_\nu(t\xi^*)
\nn\\
&\stackrel{\alpha\to 0}{\longrightarrow}&\delta(\Im m
\xi)\frac12(J_\nu(\xi)^2-J_{\nu-1}(\xi)J_{\nu+1}(\xi))\ . 
\label{rhob2w}
\eea
It agrees with the density obtained from a non-Hermitian 
one-matrix model \cite{KS}.
In the second line we indicate that in the Hermitian limit 
$\alpha \to 0$
the density
reduces to the known universal Bessel density times a delta function
in the imaginary part. In the opposite limit $\al\to\infty$ the
density at strong non-Hermiticity given in eq. (\ref{rhob2str}) can be
recovered, and we refer to \cite{AOSV} for more details. 

For $\beta=4$ the weak non-Hermiticity limit yields \cite{A05}
\bea
\rho_W^{(4,\,\Dirac)}(\xi)&=&
\frac{|\xi|^2}{32\al^4}(\xi^{\ast\,2}-\xi^2)
K_{2\nu}\left(\frac{|\xi|^2}{2\al^2}\right)
e^{\frac{(\xi^2+\xi^{*\,2})}{4\al^2}}\int_0^1 ds \int_0^1
\frac{dt}{\sqrt{t}}{e}^{-2s(1+t)\al^2} 
\nn\\
&&\times
\Big[J_{2\nu}(2\sqrt{st}\xi)J_{2\nu}(2\sqrt{s}\xi^\ast)
- J_{2\nu}(2\sqrt{s}\xi)J_{2\nu}(2\sqrt{st}\xi^\ast)
\Big].
\label{rhob4w}
\eea
Once again in the Hermitian limit $\alpha\to0$ the prefactor in front
of the integral reduces to a delta function in 
$\Im m(\xi)$, and the integral matches the known universal Bessel
kernel for $\beta=4$ (which also has a simpler representation
expressed in terms of the $\beta=2$ density plus a single integral). 

Finally we give the weak origin limit for $\beta=1$ which was obtained
most recently. The complication arises from the fact that for the real
eigenvalues both at $\mu=0$ in the real Wishart ensemble, and at
$\mu\neq0$ for our non-Hermitian extension, the large-$N$ limit of the
kernel and the integration in eq. (\ref{RrealN}) do not commute. For
more details we refer to \cite{AKPW,PhDMJP} and only quote the answer  
valid for $\xi\in\mathbb{R}$ and $\xi\in i\mathbb{R}$
\bea
\rho_{(i)\mathbb{R},W}^{(1,\,\Dirac)}(\xi) &=&
  \frac{2 |\xi|h_w(\xi^2)}{[\sgn\xi^2]^{\frac{\nu}{2}}}
  \Bigg\{\!\! \left( (-i)^{\nu} \int_{-\infty}^0 dy +
  \int_{0}^{\xi^2} \frac{2dy}{[\sgn\xi^2]^{\frac{\nu}{2}}} \right)
  {\cal K}_w(\xi^2,y) h_w(y)
 \nn \\
  && - \frac{1}{32\sqrt{\pi}}
  \Bigg[ - \, \frac{1}{\sqrt{2}\alpha}\,\e^{-2\alpha^2}\,J_{\nu}(\xi)
  + \frac{2 (\sqrt{2}\alpha)^{\nu}}{\Gamma\left(\frac{\nu+1}{2}\right)}\,
  \int_0^1 ds\,\e^{-2\alpha^2 s^2} s^{\nu+2}
  \nn
  \\
  && \times\! \left(
  \frac{\xi}{2}\,E_-(s)\,
  J_{\nu+1}(s \xi) - 2\alpha^2s\left(
  E_+(s) - E_-(s) \right)
  J_{\nu}(s \xi) \right)\!\! \Bigg]\!
  \Bigg\},
  \nn\\
&&
\eea
where we have defined the kernel in the weak limit
\be
{\cal K}_W^{(1)}(\xi,\xi^*)\equiv 
\int_0^1 \frac{ds\,s^2}{2^9 \pi\al^2}e^{-4\alpha^2 s^2}
\Big(\xi J_{\nu+1}(s\xi)J_{\nu}(s\xi^*) -  \xi^*
J_{\nu+1}(s\xi^*)J_{\nu}(s\xi)\Big), 
\label{Kb1weak}
\ee
the rescaled real weight function
\be
  h_w(\xi^2)\equiv
  \frac{2}{\xi^\nu}e^{\xi^2/16\alpha^2}\,
  2K_{\frac{\nu}{2}}\left( \frac{|\xi|^2}{16\alpha^2} \right)
\ ,
\label{hhdef}
\ee
and the exponential integral
\begin{equation}
  E_\mp(s)\equiv\int_1^\infty dt \, t^{\frac{\nu}{2}\mp
    \frac12}\,e^{-2\alpha^2s^2 t}\,. 
  \label{Edef}
\end{equation}
The weak 
density of complex eigenvalues presents no such difficulties \cite{APSoII}:
\bea
\rho_{\mathbb{C},W}^{(1,\,\Dirac)}(\xi) & =&
-16i |\xi|^2\sgn(\im \,\xi^2)e^{\frac{1}{8\al^2}(\xi^2+\xi^{*\,2})}
{\cal K}_W^{(1)}(\xi,\xi^*) \label{RcomplW}\\
&&\times
\int_0^\infty
\frac{dt}{t}e^{-\frac{t}{2^8\al^4}(\xi^4+\xi^{*\,4})-\frac{1}{4t}}
K_{\frac{\nu}{2}}\!\left(\frac{t}{2^7\al^4}|\xi|^4\right)
\erfc\Big(\frac{\sqrt{t}}{8\al^2}|\im (\xi^2)|\Big).
\nn
\eea
In the Hermitian limit $\alpha\to0$ the density of complex eigenvalues
as well as the density along the imaginary axis vanishes. Only the
density of real eigenvalues will build up the known density of real
eigenvalues in the Hermitian Wishart ensemble for $\beta=1$
\cite{APSoII}.

\section{Conclusions}
\label{open}

In this short article we have reviewed the solution of three
non-Hermitian extensions of Wishart ensembles of random matrices with
real, complex or quaternion real elements. In all three cases  
the eigenvalue correlation functions are expressed in terms of the
kernel of (skew) orthogonal Laguerre polynomials in the complex plane,
depending on a non-Hermiticity parameter that allows to interpolate
between Wishart and maximally non-Hermitian Wishart ensembles, for
finite and infinite $N$. At the origin these ensembles are very well
understood, and we gave the corresponding three Bessel kernels in the
complex plane at strong and weak non-Hermiticity. 
At the various inner and outer edges we expect that the behavior of
the corresponding Ginibre ensembles (see e.g. in \cite{BHJ}),
and the ensembles we considered here is be the same and thus universal.
This was
shown for example in \cite{ABe} for the generalized Airy kernel 
at the soft edge along the real line, or in \cite{KS} in the
rotationally invariant 
case (regime II (iii)) at the outer edge, both for $\beta=2$. The
latter behavior was also found for the outer edge density of random
contractions \cite{BHJ}.

At present a deeper guiding principle to prove
universality for non-Hermitian random matrices with a larger class of
weight functions, such as quasiharmonic potentials \cite{Zabrodin}
is lacking. A first step towards universality could be to show that
different Gaussian models lead to the same answer as mentioned above,
or as it was shown in yet 
another example for the Bessel kernel in the weak limit at $\beta=2$
eq. (\ref{rhob2w}),
starting from a Gaussian one- or two-matrix model in \cite{KJreplica} and
\cite{Osborn}, respectively. We hope that these first steps will 
ultimately lead
to a deeper understanding of the issue of universality.\\ 

{\bf Acknowledgments} I would like to thank all my coworkers for
collaborations on this subject, as well as the organizers of this
workshop for the very stimulating atmosphere.


\begin{thebibliography}{99}

\bibitem{AOSV}
G. Akemann, J.C. Osborn, K. Splittorff and J.J.M. Verbaarschot,
Nucl. Phys. {\bf B712} (2005) 287 [hep-th/0411030].

\bibitem{A05}
G. Akemann,
  Nucl. Phys. {\bf B730} (2005) 253 [hep-th/0507156].

\bibitem{AB}
G. Akemann and F. Basile,
Nucl. Phys. {\bf B766} (2007) 150
[arXiv:math-ph/0606060].

\bibitem{APSo}
G. Akemann,  M.J. Phillips and H.-J. Sommers, J. Phys. {\bf
  A}: Math. Theor. {\bf 42} (2009) 012001
[arXiv:0810.1458v1 [math-ph]].

\bibitem{APS}
G. Akemann, M.J. Phillips and L. Shifrin,
J. Math. Phys. {\bf 50} (2009) 063504 
[arXiv:0901.0897v2 [math-ph]].


\bibitem{APSoII}
  G.~Akemann, M.~J.~Phillips and H.~J.~Sommers,
  J.\ Phys.  {\bf A43} (2010) 085211
  [arXiv:0911.1276 [hep-th]].


\bibitem{ABe}
  G.~Akemann and M.~Bender,
  J.\ Math.\ Phys.\  {\bf 51} (2010) 103524
  [arXiv:1003.4222 [math-ph]].



\bibitem{AKP}
  G.~Akemann, M.~Kieburg and M.~J.~Phillips,
  J.\ Phys.  {\bf A43} (2010) 375207
  [arXiv:1005.2983 [math-ph]].


\bibitem{AKPW}
  G.~Akemann, T.~Kanazawa, M.~J.~Phillips and T.~Wettig,
  JHEP {\bf 1103} (2011) 066
  [arXiv:1012.4461 [hep-lat]].



\bibitem{BLeC}
D. Bernard and A. LeClair,
{\it A Classification of Non-Hermitian Random Matrices},
proceedings of the NATO Advanced Research Workshop on ``Statistical Field
Theories'', Como 18-23 June 2001 [arXiv:cond-mat/0110649].

\bibitem{BT}
C. Biely and S. Thurner,
Quant. Finance {\bf 8} (2008)
705 
[arXiv:physics/0609053v1 [physics.soc-ph]].


\bibitem{Burda1}
Z. Burda, A. Jarosz, G. Livan, M. A. Nowak and A. Swiech,
Phys. Rev. {\bf E82} (2010) 061114
	arXiv:1007.3594v1 [cond-mat.stat-mech].

\bibitem{Burda2}
Z. Burda, A. Jarosz, G. Livan, M. A. Nowak and A. Swiech,
arXiv:1103.3964v1 [cond-mat.stat-mech].

\bibitem{ProdBook} A. Crisanti, G. Paladin and A. Vulpiani, {\it
    Products of Random Matrices in Statistical Physics}, Springer,
  Berlin (1993).

\bibitem{Forrester07} P.J. Forrester and T. Nagao, Phys. Rev. Lett. {\bf 99}
  050603 (2007) [arXiv:0706.2020 [cond-mat.stat-mech]];
J. Phys.  {\bf A41},
375003 (2008) [arXiv:0806.0055 [math-ph]].

\bibitem{ForresterMays}
P.J. Forrester and A. Mays, J. Stat. Phys.
{\bf 134} (2009) 443
[arXiv:0809.5116v2 [math-ph]].

\bibitem{ForresterMaysII}
P.J. Forrester and A. Mays,
arXiv:0910.2531v2 [math-ph].

\bibitem{Forresterbook} P.J. Forrester, {\it Log-Gases and Random
  Matrices}, London Mathematical Society Monographs no. 34, 
Princeton University Press, Princeton (2010).


\bibitem{FKS} Y.V. Fyodorov, B.A. Khoruzhenko, and H.-J. Sommers,
Phys. Lett. {\bf A226} (1997) 46  [arXiv:cond-mat/9606173];
Phys. Rev. Lett. {\bf 79} (1997) 557  [arXiv:cond-mat/9703152].

\bibitem{FKS98} Y.V. Fyodorov, B.A. Khoruzhenko and H.-J. Sommers,
Ann. Inst. Henri Poincar\'e
{\bf 68} (1998) 449 [arXiv:chao-dyn/9802025].



\bibitem{Ginibre} J. Ginibre, J. Math. Phys. {\bf 6} (1965) 440.

\bibitem{Girko} V.L. Girko, Theor. Prob. Appl. {\bf 30} (1986) 677.


\bibitem{HOV} M.A. Halasz, J.C. Osborn and J.J.M. Verbaarschot,
Phys. Rev. {\bf D56} (1997) 7059 [hep-lat/9704007].


\bibitem{KWY}
  T.~Kanazawa, T.~Wettig and N.~Yamamoto,
  Phys.\ Rev. {\bf D81} (2010) 081701
  [arXiv:0912.4999 [hep-ph]].

\bibitem{EK} E. Kanzieper,
J. Phys. {\bf A}: Math. Gen. {\bf 35} (2002) 6631 [cond-mat/0109287].

\bibitem{KS}
E. Kanzieper and N. Singh,
	J. Math. Phys. {\bf 51} (2010) 103510
[arXiv:1006.3096v2 [math-ph]].


\bibitem{BHJ}
B.A. Khoruzhenko and H.-J. Sommers,
``Non-Hermitian Random Matrix Ensembles'',
invited chapter in ``Handbook of Random Matrix Theory'',
Eds. G. Akemann, J. Baik, P. Di Francesco, Oxford University Press 2011
[arXiv:0911.5645[math-ph]].


\bibitem{KDI-2000} J. Kwapie\'{n}, S. Dro\.{z}d\.{z}, and A.
        A. Ioannides, Phys. Rev. {\bf E62} (2000) 5557
        [arXiv:cond-mat/0002175v1 [cond-mat.stat-mech]. 

\bibitem{Kwapien}
J. Kwapie\'{n}, S. Dro\.{z}d\.{z}, A.Z. G\'orski and P. O\'swiecimka,
Acta Phys. Pol. {\bf B37} (2006) 3039 
[arXiv:physics/0605115v1 [physics.soc-ph]].



\bibitem{Magnea}
U. Magnea,
J. Phys.  {\bf A41} (2008) 045203 [arXiv:0707.0418v2 [math-ph]].

\bibitem{MP} V.A. Marchenko and L.A. Pastur, Math. USSR-Sb, {\bf 1} (1967)
457.


\bibitem{Osborn} J.C. Osborn,
Phys. Rev. Lett. {\bf 93} (2004) 222001 [hep-th/0403131].

\bibitem{Karol} 
K.A. Penson, K. \.Zyczkowski, 	arXiv:1103.3453v2 [math-ph].

\bibitem{PhDMJP}
M.J.~Phillips, {\it }
PhD thesis, Brunel University West London, 2011.


\bibitem{Sommers2008}
H.-J. Sommers and W. Wieczorek,  J. Phys.  {\bf A41} (2008) 405003
[arXiv:0806.2756 [cond-mat.stat-mech]].



\bibitem{KJreplica} K. Splittorff and J.J.M. Verbaarschot,
Nucl. Phys. {\bf B683} (2004) 467 
[arXiv:hep-th/0310271v3].

\bibitem{Steph} M. Stephanov, { Phys. Rev. Lett.} {\bf 76} (1996) 4472
[hep-lat/9604003].


\bibitem{Jac}
  J.J.M.~Verbaarschot,
  ``Handbook Article on Applications of Random Matrix Theory to QCD,''
invited chapter in "Handbook of Random Matrix Theory",
Eds. G. Akemann, J. Baik, P. Di Francesco, Oxford University Press
2011 
  [arXiv:0910.4134 [hep-th]].


\bibitem{Zabrodin} A. Zabrodin, 
"Random matrices and Laplacian growth", invited chapter in "Handbook
of Random Matrix Theory", Eds. G. Akemann, J. Baik, P. Di Francesco,
Oxford University Press 2011 
[arXiv:0907.4929v1 [math-ph]].


\end{thebibliography}
\end{document}